# CredSaT: Credibility Ranking of Users in Big Social Data incorporating Semantic Analysis and Temporal Factor

**Bilal Abu-Salih, Pornpit Wongthongtham, Kit Yan Chan, Dengya Zhu**

**Curtin University**
**Australia**

**Abstract**

The widespread use of big social data has pointed the research community in several significant directions. In particular, the notion of social trust has attracted a great deal of attention from information processors / computer scientists and information consumers / formal organizations. This is evident in various applications such as recommendation systems, viral marketing and expertise retrieval. Hence, it is essential to have frameworks that can temporally measure users' credibility in all domains categorised under big social data. This paper presents CredSaT (**Cred**ibility incorporating **S**emantic **a**nalysis and **T**emporal factor): a fine-grained users' credibility analysis framework for big social data. A novel metric that includes both new and current features, as well as the temporal factor, is harnessed to establish the credibility ranking of users. Experiments on real-world dataset demonstrate the effectiveness and applicability of our model to indicate highly domain-based trustworthy users. Further, CredSaT shows the capacity in capturing spammers and other anomalous users.

***Keywords:*** *Domain-based Credibility, Big Social Data, Information Retrieval, Semantic Analysis, Temporal Factor.*

## 1. INTRODUCTION

Deep insights into Big Data (BD) require a better understanding of the massive amount of data being generated every second, necessitating the leveraging of new data analysis techniques and the continuous improvement of existing practices. Researchers are trying to capture the Value of BD in dissimilar contexts. In Online Social Networks (OSNs), it is important to understand the users' behaviour because of the dramatic increase in the usage of online social platforms. This explains the importance of measuring the users' trustworthiness, and ascertaining the users' influence in a particular domain. For example, "many marketing researchers believe that social media analytics presents a unique opportunity for businesses to treat the market as a 'conversation' between businesses and customers" [1]. Hence, the factual grasp of the users' domains of interest and an appropriate judgement of their emotions enhances the customer-to-business engagement. This necessitates an accurate analysis of customer reviews and their opinions in order to improve brand loyalty, improve customer service, and increase an organization's awareness of issues that need to be addressed. The incorporation of semantic analysis in OSNs, in particular, reduces the ambiguity of Big Social Data (BSD) by clarifying the actual context of the users' content. This mitigates the variability of BD [2] [3], distinguishes users' domains of interest, and deduces their actual sentiments.

Sentiment analysis (a.k.a. opinion mining) has become a core dimension of researchers' endeavours to create applications that leverage the massive increase of user-generated content [4]. For example, in OSNs, sentiment analysis has been utilized in several aspects of research [5-7]. In the context of social trust, frameworks have been developed to analyse the trustworthiness of users' content, taking into consideration the overall feelings regarding what they have chosen to expose in their content. However, in terms of trustworthiness analysis, these efforts did not attempt a sentiment analysis of a post's replies. Likewise, most of these efforts assimilated the sentiment analysis of the content regardless of its context. Hence, the semantic analysis should be amalgamated to improve the resultant sentiment.

The *veracity* of BD [8] refers to the accuracy, correctness and trustworthiness of data. Demchenko et al. [8] presented multiple factors to ensure the *veracity* of BD. These factors include, but are not limited to, the following: (i) trustworthiness of data origin; (ii) reliability and security of data store; (iii) data availability. This list could be enhanced by including a further two essential aspects: *correctness* and *consistency*. Although data origin and store are critical, the trustworthiness of the source does not guarantee data correctness and consistency. Data cleansing and integration should be incorporated to ensure the *veracity* of data as well. Further techniques are implemented in our approach to achieve the *veracity* of data which is significant for further BD analysis.

One of the main reasons for acquiring the *value* of BSD is to provide frameworks and methodologies by means of which the credibility of OSNs users can be evaluated. These approaches should be scalable to accommodate large-scale BSD. In our previous work [9] we proposed a preliminary approach to measure the domain-based trustworthiness of OSNs users. In this work, we present CredSaT (***Cre**dibility incorporating **S**emantic **a**nalysis and **T**emporal factor*): a comprehensive BSD framework to measure users' credibility in all domains. The crawled datasets of user data and metadata are divided into several chunks where each chunk represents a specific time period. A metric of key credibility attributes is incorporated to evaluate the user's credibility in each particular chuck, thus providing an overall credibility value. The mechanism used to calculate a user's value in each step takes into account the values of other users, thereby providing a normalization approach for establishing a ranking list of credibility in each domain.

To evaluate the applicability and effectiveness of CredSaT, we benchmark our approach against prior state-of-art approaches in identifying highly domain-based trustworthy users. Additionally, further experiments conducted to validate the proposed approach in capturing low trustworthy users.

The main contributions of this paper are summarised as follows:
- CredSaT is proposed as an effective credibility framework for users of OSNs addressing the main features of BD, and incorporating semantic analysis and the temporal factor.
- A novel metric incorporating key attributes is incorporated to measure the domain-based credibility of users.
- The evaluation of our approach verifies its effectiveness as it is capable of discovering influential domain-based users, and users with anomalous behaviour.

The rest of this paper is organized as follows: Section II reviews other work related to the issue of credibility in OSNs. A methodology presenting the framework of the proposed approach is described in Section III. Section IV presents the data generation, acquisition and the pre-processing phases. Data storage and analysis, including the core modules, are illustrated in Section V. Section VI presents the conducted experiments to validate our approach. Section VII addresses certain limitations of the proposed framework and mentions the anticipated improvements that will be made by future work.

## 2. RELATED WORK

There have been several efforts to define and formulate the trustworthiness of users in OSNs. Although most of these efforts presented generic-based credibility evaluation approaches [10-12], subsequent studies have focused on the users' topic of interest. Zhao et al. [13] proposed a scalable trustworthiness inference module for twitterers and their tweets that takes into account the heterogeneous contextual properties. Another host of scholars have addressed the issue of influential users in OSNs [14-19]. One such well-known effort to discover influential users is that of Twitterrank [20]; however, the main limitations of their approach are the omission of a temporal factor, and the use of the bag-of-words technique which disregards the semantic relationships of terms in a document. Understanding the users' behaviour over time and addressing their interests by incorporating semantic analysis is a significant extension of these efforts. In this context, our previous work [9] highlighted the notion of using trust for the data extracted from unstructured content (such as social media data) in order to calculate trustworthiness values which correspond to a particular user in a particular domain. The literature of trust in social media shows a lack of methodologies for measuring domain-based trust. Ontology represents the core of the domain where the knowledge is shared amongst different entities within the system that may include people or software agents [21].

The use of sentiment analysis techniques to analyse the content of OSNs, has significantly influenced several aspects of research. In the context of social trust, authors of [22] proposed a recommendation system framework incorporating implicit trust between users and their emotions. AlRubaian et al. [23] presented a multi-stage credibility framework for the assessment of microbloggers' content. The development of sentiment-based trustworthiness approaches for OSNs is discussed further in [5-7]. However, in their trustworthiness analysis, these efforts did not analyse the sentiment of a post's replies. Furthermore, the sentiment analysis of the content was conducted regardless of its context. A bag-of-words topic inference technique such as Latent Dirichelet Allocation (LDA) is not appropriate for classifying short text like tweets by high-level topics [24]. Hence, semantic analysis should be included in order to improve the resultant sentiments and better understand the context of the text.

### 3. CREDSAT FRAMEWORK

Figure 1 illustrates the CredSaT framework. This framework adopts the BD value chain presented by Hu et al. [25] which covers the life cycle of BD. This chain comprises four main stages: data generation, data acquisition, data storage, and data analysis.

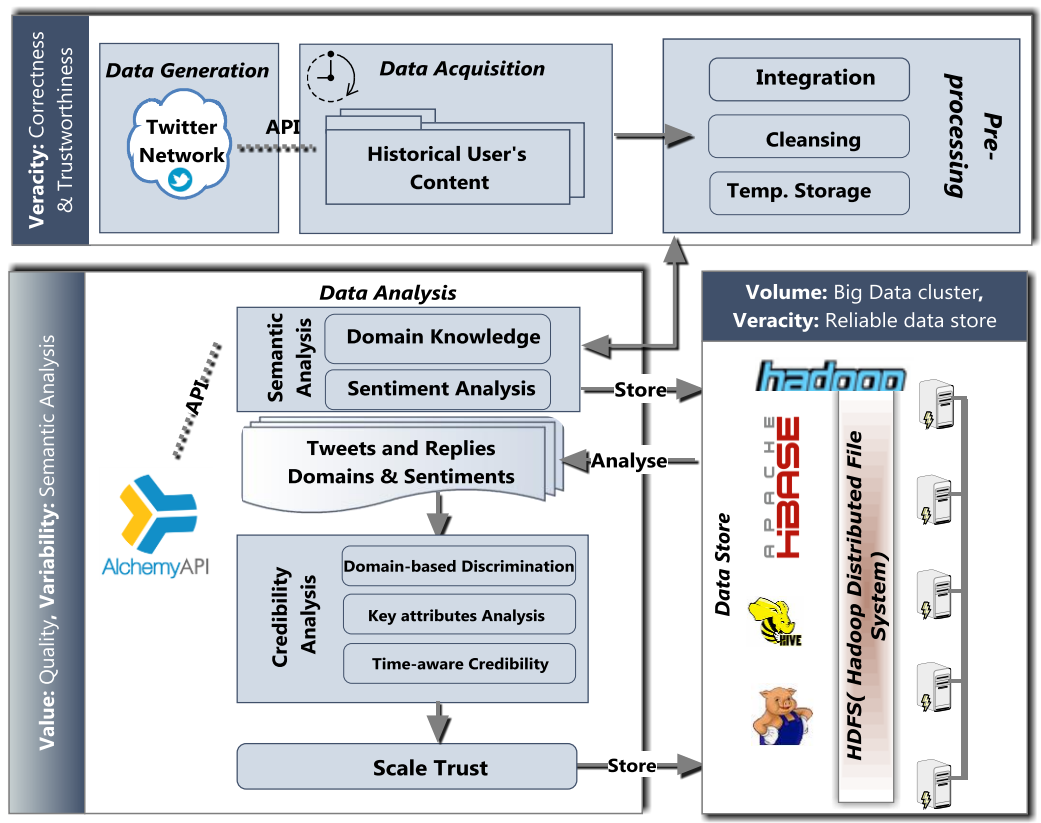

Fig. 1. CredSaT Framework

## 4. DATA GENERATION AND ACQUISITION

### 4.1 Data Generation

Twitter micro-blogging has been extensively utilized to conduct research in several domains of interest. This paper focuses on the data generated from Twitter; however, the proposed approach is applicable to all other social media platforms.

### 4.2 Data Acquisition

We have collected the data for this study using Twitter data access mechanisms. Users' information and their tweets and all related metadata were crawled using *TwitterAPI* [26]. Data acquisition is carried out by means of a PHP script triggered by running a cron job which selects a new user_id and starts collecting historical user information, tweets, replies and the related metadata. The list of twitterers' user_ids used in the data acquisition phase is extracted from a Twitter graph dataset crawled by Akcora et al. [27]. This graph is chosen since it includes the list of users who had less than 5,000 friends in 2013. This threshold was established by Akcora et al. to discover bots, spammers and robot accounts. We used this threshold to measure their credibility as well. This assists in finding domain influencers from a dataset of general users whose domains of knowledge are not explicitly known. Further, our framework is capable of identifying anomalous users as illustrated in the evaluation section.

Fig. 2 shows the number of crawled tweets posted between 2006 and 2015. As can be seen in Fig. 2, there has been a dramatic increase in the number of tweets since 2006. This indicates the great extent to which people are engaging with social media. These social platforms enable them to publish their content, taking advantage of the open environment and fewer restrictions.

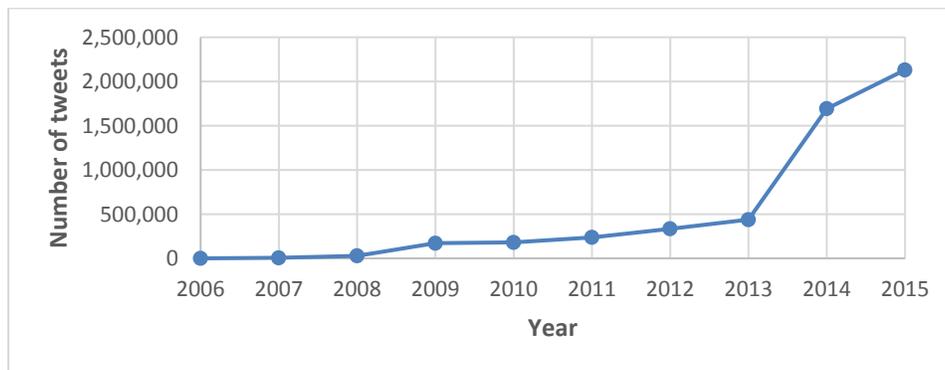

Fig. 2. Number of tweets per year

### 4.3 Pre-processing phase

To address the data veracity in terms of data correctness, the raw extracted tweets were subjected to a pre-processing phase. This phase includes the following steps:

*Data integration and temporary storage:*

The raw extracted tweets from TwitterAPI are in JSON format. *AcTwitterConversation* fetches conversation tweets as an array. Data integration will provide a better structured data format for the next analysis phase. This has been achieved by reformatting and unifying the raw tweets (JSON) and replies (ARRAY) to fit with the database relational model, the design of which is based on the metadata of the tweet, reply and the user. Then, the reformatted data is stored in a temporary location (i.e. MySql database).

*Data cleansing:*

Data at this stage may include many errors, meaningless data, irrelevant data, redundant data, etc. Thus, data is cleansed to remove noisy data and ensure data consistency. The following sequential steps were taken for the data cleansing process (i) we eliminated all redundant content; (ii) users who posted fewer than 50 tweets were excluded. We established this particular threshold because the aim of this research is to discover domain-based influential users; thus, we assume that those users post a relatively large number of tweets in their domain(s) of interest; (iii) we eliminated media *urls* such as photos uploaded to twitter, or media uploaded to one of the popular media sharing websites listed in [28] such as Instagram, Flickr, YouTube, and Pinterest. This is because these have no actual text that can be extracted for further analysis.

## 5. DATA STORAGE AND ANALYSIS

Data storage is the third phase of the BD lifecycle. *Volume* is an essential dimension to be considered when describing BD. The BD infrastructure at the School of Information Systems, Curtin University, is utilized for data storage. The temporal-temporary data dumps its contents to this distributed environment after the data integration process. Although the size of our data could be stored and managed using one computer, the BD cluster is utilized as an infrastructure required for our continuous research in BD analysis incorporating large scale, heterogeneous types of data.

The data analysis phase is the production stage of the BD value chain. The credibility of users, incorporating semantic dimension and the temporal factor, is established as explained in the following sub-sections.

### 5.1 Domain Extraction and Sentiment analysis

Deep insights into Big Data (BD) require new data analysis techniques and the continuous improvement of existing practices. This mitigates the variability of BD [2, 3], distinguishes users' domains of interest and infers their genuine sentiments.

In this context, AlchemyAPI[1] is used as a domain knowledge inference tool to infer the content's taxonomies. AlchemyAPI analyses the given text or URL and categorizes the content of the text or webpage according to three domains (taxonomies) with the corresponding *scores* and *confident* values. *Scores* are calculated using AlchemyAPI, range from "0" to "1", and convey the correctness degree of an assigned Taxonomy/Domain to the processed text or webpage. *Confident* is a flag associated with each response, indicating whether AlchemyAPI is confident with the output. AlchemyAPI is used further to identify the overall positive or negative sentiment of the provided content.

A tweet's content has one or two main components: *text* and *url*. Due to the limitation of a tweet's length, a normal or legitimate twitterer attaches with her tweet a URL to a particular webpage, photo, or video to help her followers obtain further information on the tweet's topic. Twitter scans URLs against a list of potentially harmful websites, then URLs are shortened using *t.co service* to maximise the use of the tweet's length. Anomalous users such as spammers abuse this feature by hijacking trends, using unsolicited mentions, etc., to attach misleading URLs to their tweets. Thus, it is important to study the tweet's domain and the comprised URL's domain to obtain a better understanding of the user's domain(s) of knowledge, which are then used to measure user domain-based credibility.

AlchemyAPI is used to analyse and infer taxonomies of each user's tweet and the website content of the associated URL rather than analysing the user's timeline as one block. This is in order to provide a fine-grained analysis of tweet data. AlchemyAPI may not able to infer a domain for any particular tweet or URL when the tweet is very short, or the content is unclear or nonsensical, or written in a language other than English. Likewise, if the URL is invalid, corrupted or contains non-English content, domains cannot be inferred. Currently, English language contents are the only contents supported by AlchemyAPI in their taxonomy inference technique. Hence, we remove a tweet and its metadata from the dataset if the tweet was written in another language.

We utilize AlchemyAPI further in this paper to derive the sentiment of a given reply whether it is positive, negative or natural with the corresponding sentiment score. Consequently, all of a tweet's set of replies are crawled and the sentiments of these replies are incorporated in the analysis to enhance the credibility process as discussed later in Section 5.2.3.

Table 1: Total count of users, tweets and replies before and after cleansing phase

|  | Before Cleansing | After Advanced Cleansing | Eliminated (%) |
|---|---|---|---|
| *Total # Users* | 9,772 | 7,401 | 24.3% |
| *Total # Tweets* | 5,220,478 | 2,810,362 | 46.2% |
| *Total # Replies* | 2,010,992 | 1,443,932 | 28.2% |

---

[1] http://www.alchemyapi.com/

Table 1 shows the total number of users, tweets and replies before and after the advanced cleansing process. It is worth noting the importance of data cleansing to purify the raw dataset and enhance its quality. Although the selection criteria for the OSNs' users in this research are quite restricted, the number of eliminated content highlights some significant issues as follows: (i) the quality of contents posted in the social media should be critically studied before conducting further analysis; (ii) it is important in the cleansing phase to ensure the data *veracity* in the BD context; (iii) part of the eliminated tweets were written in a non-English language; however, since these tweets may contain valuable content, sophisticated semantic analysis tools are required to address multilingual contents.

Fig. 3 displays the list of all domains and corresponding numbers of users, tweets and URLs extracted from the dataset after the cleansing process. It is evident that "arts and entertainment" is the area of most interest to tweeters. For example, 4,550 users tweeted about 'art and entertainment' domain, and the total number of tweets in this category is 471,883. On the other hand, "real estate" seems to be the area of less interest to users; 2,596 and 9,163 are the total numbers of users and tweets respectively. As depicted in Figure 3, the number of users in each domain is relatively high. This is because AlchemyAPI inferred a wide range of domains from users' content. For example, 32% of the users posted at least one tweet or URL for each domain. We will discuss this issue further in the credibility analysis phase.

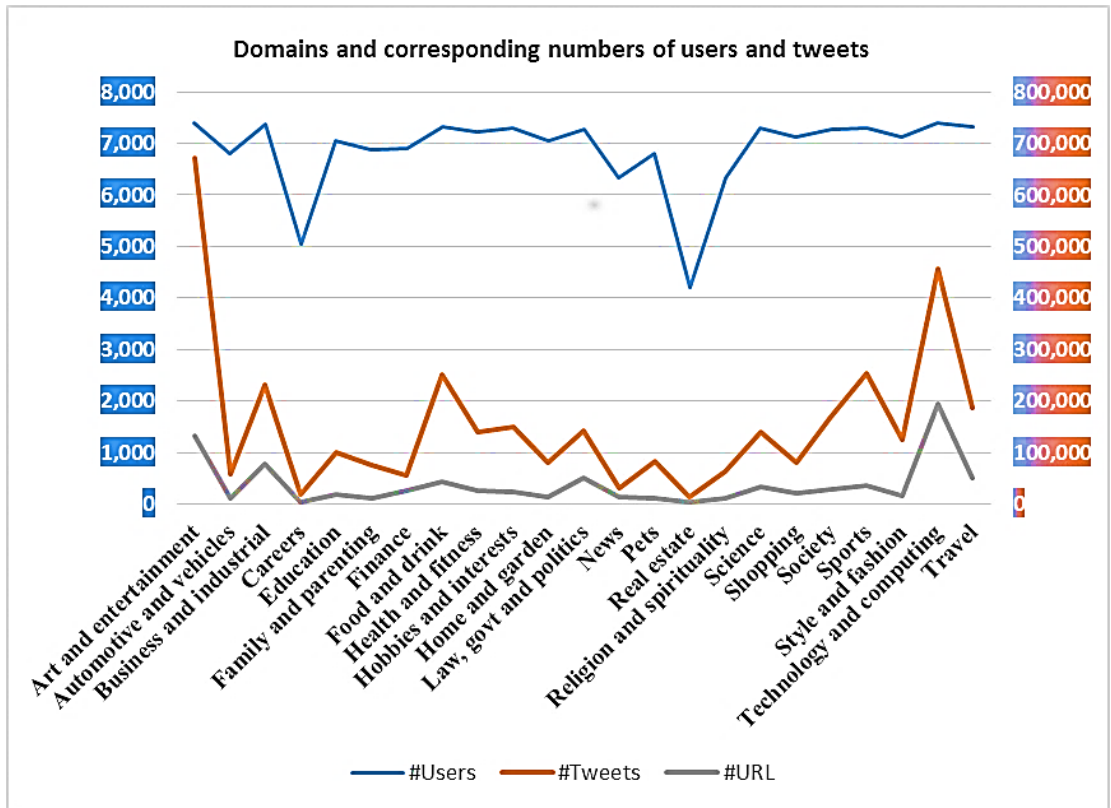

Fig. 3. Total Numbers of users, tweets and urls in each taxonomy

## 5.2 Time-aware domain-based user's credibility analysis

A domain-based analysis of users' credibility is suggested in order to provide a comprehensive scalable framework. This is achieved by analysing the collection of a user's tweets in order to measure the initial user's credibility value based on the user's historical data. This is done through the time-aware, domain-based user credibility ranking approach.

### 5.2.1 Methodology

One of the main objectives of CredSaT is to discover influential domain-based users from the list of users whose domain(s) of knowledge is tacit, incorporating the temporal factor. Hence, the outcome should be a ranked list of users with a corresponding credibility value for each specific domain. The mechanism used to calculate a user's value in each step takes into account other users' values, thereby providing a normalization approach for building the relative ranking list of credibility in each domain. Hence, each of the key attributes is normalized in each domain by dividing the value of the user's attribute by the maximum value achieved by all users in that domain. The following subsections explain the metric used to measure the domain-based credibility of users in BSD.

### 5.2.2 Distinguishing domain-based OSNs users

The analysis of a user's content for the purpose of discovering his/her main domains of interest is an essential start to the process of measuring the user's credibility. In OSNs, a user $u$ achieves a higher weight value in a certain domain(s) of knowledge if $u$ shows a strong interest in these domains through the posted tweets and attached URLs. This weight should be higher than those of other users who posted content in a broad range of domains. This is due to the fact that no user could be conversant with all domains of knowledge [29]. Therefore, the theoretical notion of Term Frequency-Inverse Document Frequency (TF-IDF) has been used to distinguish domain-based users of OSNs from others [9]. This involves studying the content of users' tweets and their embedded URLs in order to obtain a thorough understanding of their domain(s) of knowledge. To this end, a domain-based user's content score matrix is proposed:

**Domain-based User's Content Scores Matrix (*Sc*):** $Sc_{u,d}$ refers to the refinement summing of the corresponding scores achieved by AlchemyAPI for all tweets' texts $(Sc_{u,d}^{Twt})$, and the refinement summing of scores retrieved from URLs' webpage content $(Sc_{u,d}^{URL})$ posted by a user $u$ where a domain $d$ was inferred. It can be calculated as follows:

$$Sc_{u,d} = (Twt\_Sim_u \times Sc_{u,d}^{Twt} + URL\_Sim_u \times Sc_{u,d}^{URL}), \text{for each domain } d \quad (1)$$

where $Sc_{u,d}^{Twt}$ is computed by adding all scores retrieved from AlchemyAPI of tweets' texts posted by user $u$ in domain $d$, $Sc_{u,d}^{URL}$ is calculated by accumulating scores for all websites' content of the URLs embedded in users $u$'s tweets in domain $d$, $Twt\_Sim_u$ is the *Tweets Similarity Penalty* factor, it can be defined as follows:

***User's Tweets Similarity Penalty (Twt_Sim):*** where $Twt\_Sim_u$ represents the count of unique keywords (***#DistinctWords***) in the overall user's tweets to the total number of keywords in the user's tweet (***#Words***). $Twt\_Sim_u$ can be calculated as:

$$Twt\_Sim_u = \frac{\#DistinctWords_u}{\#Words_u} \quad (2)$$

$URL\_Sim_u$ in Eq. (1) is the *URL Similarity Penalty* factor, and is defined as follows:

***User's URL Similarity Penalty (URL_Sim):*** where $URL\_Sim_u$ represents the percentage of non-redundant URLs (***#DistinctURLs***) with non-redundant hosts of URLs (***#DistinctURLsHosts***) to the total number of URLs (***#URLs***) posted by user ***u;*** it is computed as follows**:**

$$URL\_Sim_u = 0.5 \left( \frac{\#DistinctURLs_u + \#DistinctURLsHosts_u}{\#URLs_u} \right) \quad (3)$$

$\#DistinctURLs_u$, $\#DistinctURLsHosts_u$, and $\#URLs_u$ could have the same value; i.e. the user might add a unique *url* for each time she attaches a *url* to a tweet. Thus, "0.5" is added to normalize $URL\_Sim_u$ value. Table 2 shows a list of synthetic tweets to illustrate the idea of similarities penalty. As illustrated in Table 2, only the highlighted words are counted in calculating the similarity of tweet texts $Twt\_Sim_x$. The $Twt\_Sim_x$ is computed after eliminating the stopwords and URLs from the tweets text. $URL\_Sim_x$ is calculated by extracting all urls, and finding the non-redundant URLs and hosts.

Table 2: Synthetic Tweets to illustrate the use of URL and Tweets' texts similarities.

| **List of Tweets** | *Tweet1:* "This **website** is **amazing** and **useful**: http://www.example.com/subdirectory1/index.html" *Tweet2:* "**Check** this **website** for **recent update**: http://www.example.com/index.html" *Tweet3:* "**Check** this **website** for **update**: http://www.example.com/subdirectory2/index.html" | | | | | | | | |
|---|---|---|---|---|---|---|---|---|---|
| ***Words→Count*** | 'website' → 3, 'amazing'→1, 'useful'→1, 'check'→ 2, 'recent'→1, 'update'→2 | | | | | | | | |
| ***Distinct Words*** | 'website', 'amazing', 'useful', 'check', 'recent', 'update' | | | | | | | | |
| ***DistinctURLs*** | 'http:// www.example.com /subdirectory1/index.html' 'http://www.example.com/index.html' 'http://www.example.com/subdirectory2/index.html' | | | | | | | | |
| ***DistinctURLs Hosts*** | www.example.com | | | | | | | | |
| *#Distinct Words* | 6 | *#Words* | 10 | $Twt\_Sim_x$ | 0.6 | *#Distinct URLs* | 3 | *#Distinct URLs Hosts* | 1 | *#URLs* | 3 | $URL\_Sim_x$ | 0.666 |

*URLs similarity penalty* and *Tweets similarity penalty* are proposed to address the similarities of embedded URLs and texts of the users' tweets. Generally, legitimate (normal) users who are knowledgeable or influencers in a certain domain(s) do not post the same tweet(s) repeatedly[30], or embed in their tweets in the same URL(s).

In this paper, our intention is to distinguish between the knowledgeable domain-based users and users engaged in anomalous behavior. Therefore, the aforementioned user categories should be assigned less weight than the normal user category; hence, $Twt\_Sim_u$ and $URL\_Sim_u$ are proposed as penalty factors applied to the scores given to the text of the user's tweet or content of embedded URLs' websites.

AlchemyAPI infers a maximum of three different taxonomies for each text or webpage; however, the corresponding scores are considered as an important factor. Taxonomy with a score of value '1' should acquire a higher weight than taxonomy with a score of value '0.2'. Thus, $Sc_{u,d}$ is proposed which accumulates for user $u$ the refinement resultant scores of domain $d$ from the list of historical user's tweets and the websites' content of the attached URLs.

It is important to obtain an overall understanding of the domain(s) in which each user is interested; hence, domain frequency ($DF_u$) is incorporated which calculates the number of domains the user $u$ has tweeted about. To distinguish users among the list of their domains of interest, the inverse domain frequency set (**IDF**) is implemented as follows:

$$IDF_u = log(\frac{n}{DF_u}) \qquad (4)$$

*Where* $n$ = the number of domains, $DF_u$ = the domain frequency for user $u$.

$IDF_u$ is proposed to assign the user a high weight if her content and embedded URLs are indicating only a few domains. On the other hand, $IDF_u$ assigns a low weight to a user if the content of her tweets and URLs are indicating a large range of domain(s).

The last step of this phase is used to calculate the weight for each user in each domain by combining the following factors: (i) domain-based user's content scores **Sc** (users' interest in each domain); (ii) the normalized inverse domain frequency **IDF** (distinguish users amongst domains of interest) as follows:

$$W_{u,d} = Sc_{u,d} \times IDF_u, \qquad where\ Sc_{u,d} > \rho \qquad ,\text{for each domain } d \qquad (5)$$

where $W_{u,d}$ represents the weight of each user $u$ obtained in the domain $d$, $\rho$ is a threshold value provided as a fine-tuning parameter representing the minimum total scores for each user in each domain. The imposition of this threshold is intended to provide more accurate and reasonable results. In particular, having this threshold means that the small **Sc** scores achieved by each user in each domain will be disregarded. This is because a small user's scores in each domain could end up decreasing the overall discriminating value of this user in all domains. Users may deviate from their domain of knowledge to discuss general, unrelated or trending topics.

Entries of **W** are normalized into the range of [0, 1] by dividing each entry by the maximum weight values of the corresponding domain. For example, all users' weight values in the domain $d$ are normalized as follows:

$$W'_{u,d} = \frac{W_{u,d}}{max(W_{*d})} \quad , \text{for each domain } d \tag{6}$$

where $W'_{u,d}$ is the normalized weight of user $u$ in domain $d$, $max(W_{*d})$ represents the maximum weight value of all collected users in domain $d$.

Although the weight assigned to each user in each domain is important to address the interests of OSNs' users; however, this is should be consolidated through the analysis of users' metadata. This will obtain a comprehensive insight into the users' behaviour based on their interactions with other users in the OSN platform. This dimension will be addressed in the next section

### 5.2.3 Feature-based user ranking

It is important to have an understanding of the interactions-based attributes of OSN users, as this is a significant factor when discovering socially reliable, domain-based users. This involves studying the followers' interest in the users' content, their positive or negative opinions, etc. In this section, a metric incorporating several key attributes is used to build the feature-based ranking model.

As mentioned previously, AlchemyAPI infers a maximum of three taxonomies for each processed text (i.e. tweet's text or URL's website content). The tweets' metadata (such as #likes, #Retweet, #Replies, etc.) does not indicate the particular domain in which the follower has valued the tweet. Hence, the user's scores produced by AlchemyAPI for each domain are used to provide a weighting distribution mechanism for all metadata items in the inferred domains; we termed this mechanism by *domain-base relativeness factor*. More details will be provided under each feature in the following subsections.

***Domain-based user retweet matrix*** ($R$), where $R_{u,d}$ represents the frequency of retweets for user' content in each domain $d$.

The *domain-base relativeness factor* is used to calculate $R_u$ based on the $u$'s score obtained for each domain $d$. In particular, total count of retweets "*retweet_count*" is distributed among the $u$'s domain(s) based on her score for each one. For example, suppose the domain-base scores spreading for a tweet ($t_x$) posted by user $u$ is (1, 0.5, and 0.5) in ("Sports", "Arts and Entertainment", and "Education") domains respectively, and the total retweets of $u$'s tweet = **10**, then the distribution number of retweets for user $u$ is ($R_{u,sports}$ = 5, $R_{u,arts}$ = 2.5, $R_{u,education}$ = 2.5). $R$ is normalized as follows:

$$R'_{u,d} = \frac{R_{u,d}}{max(R_{*d})} \quad , \text{for each domain } d \tag{7}$$

Where $max(R_{*d})$ is the maximum count of retweets obtained for all users' content in domain $d$.

It is evident that the crawled dataset for any user might contain one or more of the following categories: original tweets, retweets or replies to other tweets. The content of retweets has been retained and used for domain discovery purposes. When a user retweets a certain tweet $t_y$ then she supports the context of $t_y$ despite $t_y$ being originated by someone else. However, all retweets with the associated metadata have been eliminated and are not counted for credibility purposes. This is because the metadata such as (retweet_count, favorite_count, and replies_count) which are associated with this tweet's category indicate the original tweet and cannot be used to support the credibility of the retwitterer.

***Domain-based user likes matrix*** (***L***), where $L_{u,d}$ represents the percentage of likes/Favourites count for the users' content in each domain ***d***. ***L*** is normalized as follows:

$$L'_{u,d} = \frac{L_{u,d}}{max(L_{*d})} \quad \text{,for each domain } d \quad (8)$$

Where **max**($L_{*d}$) is the maximum percentage of likes/Favourites obtained for all users' content in domain ***d***. "*fav_count*" metadata value is distributed based on the *domain-base relativeness factor* mechanism.

***Domain-based user replies matrix*** (***P***), where $P_{u,d}$ embodies the count of replies to the users' content in each domain ***d***. ***P*** is normalized as follows:

$$P'_{u,d} = \frac{P_{u,d}}{max(P_{*d})} \quad \text{,for each domain } d \quad (9)$$

Where **max**($P_{*d}$) is the maximum percentage of replies obtained for all users' contents in domain ***d***. "*replies_count*" metadata is distributed based on *domain-base relativeness factor* mechanism. Still, the domains associated with the content of tweets' replies can be analysed to extract the actual domain(s) of each replies. This will be addressed in our future research in order to enhance the entries of ***P***.

In OSNs, sentiment analysis has been utilized in several aspects of research. In the context of social trust, frameworks have been developed to analyse the trustworthiness of users' content taking into consideration the overall feelings towards what users expose in their content. However, these efforts did not analyse the sentiment in a post's replies in evaluating trustworthiness of users and their content The following are the features proposed to address the analysis of replies in terms of sentiment.

***Domain-based user positive sentiment replies matrix*** (***SP***), where $SP_{u,d}$ refers to the sum of the positive scores of all replies to a user ***u*** in domain ***d***. Positive scores are achieved from AlchemyAPI with values greater than "0" and less than or equal to "1". The higher the positive score, the greater is positive attitude the repliers have to the users' content.

***Domain-based user negative sentiment replies matrix*** (***SN***), where $SN_{u,d}$ represents the sum of the negative scores of all replies to a user ***u*** in domain ***d***. Negative scores are those values greater than or equal to "-1" and less than "0". The lower the negative score, the greater is the negative attitude the repliers have to the users' content.

***Domain-based user sentiments replies matrix*** $(S)$, where $S_{u,d}$ embodies the difference between the positive and negative sentiments of all replies to user $u$ in the domain $d$. $S$ is normalized as follows:

$$S'_{u,d} = \frac{S_{u,d} - min(S_{*d})}{max(S_{*d}) - min(S_{*d})} \text{ , where } \quad S_{u,d} = SP_{u,d} - |SN_{u,d}| \text{ , for each domain } d \quad (10)$$

$S_{u,d}$ embodies the difference between the positive scores and the negative scores for all replies to user $u$ in domain $d$. $max(S_{*d})$ represents the maximum differences between the positive and negative replies to all collected users in domain $d$. $min(S_{*d})$ represents the minimum differences between the positive and negative replies to all collected users in domain $d$. It is evident that the list of replies could include responses from the tweet's owner as a part of the conversation. All replies posted by the tweet's owner are eliminated from the conversation and are not included in the above equations. This is in order to provide accurate sentiments results which reflect the actual positive or negative opinions of the tweet expressed by its followers. The entries of $SP$ and $SN$ are computed using the *domain-base relativeness factor* mechanism. For example, suppose *replies_count* for the tweet $(t_x)$ of the example mentioned before is equal to **10**, and the sum of the positive and negative replies for $t_x$ are (**15, -10**) respectively, then the dispersal of the positive scores amongst the extracted domains will be **($SP_{u,sports}$ = 7.5, $SP_{u,arts}$ = 3.75, $SP_{u,education}$ = 3.75),** and the dispersal of the negative scores is **($SN_{u,sports}$ = -5, $SN_{u,arts}$ = -2.5, $SN_{u,education}$ = -2.5).**

The last dimension from the list of user's key attributes is the relationship between the number of followers and friends of each user. This relation has been incorporated in the literature to measure the credibility of the OSNs' users; Wang [30] used this relation to provide a reputation measurement for the user. This measurement tool is improved in this paper as follows:

***User Followers-Friends Relation matrix*** $(FF\_R)$, where $FF\_R_u$ refers to the difference between the number of followers and friends that user $u$ obtains to the *age* of user's profile. $FF\_R_u$ is calculated as follows:

$$FF\_R_u = \begin{cases} \frac{FOL_u - FRD_u}{Age_u}, & \text{if } FOL_u - FRD_u \neq 0 \\ \frac{1}{Age_u}, & \text{if } FOL_u - FRD_u = 0 \end{cases} \quad (11)$$

Where $FOL_u$ is the number of $u$'s followers, $FRD_u$ is the number of $u$'s friends, and $Age_u$ is the age of $u$'s profile in years. The variance between the numbers of followers and friends could be due to the profile's age. Users who obtained a dramatic positive difference between number of followers and friends during a relatively short period have an advantage over those who have achieved the same difference albeit over a long period of time. $FF\_R_u$ is normalised as follows:

$$FF\_R'_u = \frac{FF\_R_u - min(FF\_R)}{max(FF\_R) - min(FF\_R)} \tag{12}$$

Where **max(FOL)** is the maximum *Followers-Friends Ratio* value of all users in the network, **min(FRD)** is the minimum *Followers-Friends Ratio* value of all users in the network.

### 5.2.4 Monitoring user's credibility over time

So far, we have presented the list of features used to measure the credibility of OSNs users. It is evident that the individual preliminary results of each key attribute cannot be used to judge the credibility of the user as such; to ascertain credibility, all available data and metadata should be analyzed thoroughly in order to produce an accurate measurement of the trustworthiness of users. An initial holistic domain-based user credibility formula incorporating all key attributes is proposed as follows:

$$C_{u,d} = \alpha * FF\_R'_u + \beta * W'_{u,d} + \gamma * R'_{u,d} + \delta * L'_{u,d} + \theta * P'_{u,d} + \vartheta * S'_{u,d} \tag{13}$$

where $C_{u,d}$ represents the user **u**'s credibility in domain **d**, while ($\alpha, \beta, \gamma, \delta, \theta, \vartheta$) are introduced to adjust the significance of each key attribute (*where $\alpha + \beta + \gamma + \delta + \theta + \vartheta = 1$*). Although $C_{u,d}$ provides a broad view of the user's trustworthiness in each domain of knowledge, the temporal factor is necessary to observe the user behavior over time thus consolidate the proposed approach. The temporal factor is assimilated as follows:

$$TC_{u,d} = \frac{\sum_{k=1}^{I} w(k) \times C_{u,d}^{k}}{\sum_{k=1}^{I} w(k)} \tag{14}$$

where $TC_{u,d}$ is the new time-aware domain-based credibility of user **u** in domain **d**, $C^k$ is the domain-based credibility matrix which is calculated for the time period **k**, and $w(k)$ is a weighting function introduced to provide a weighting mechanism for each credibility value of each time period. **I** is a *credibility window* defined as follows:

**Definition 1.** C*redibility Window (I)*: is the number of the twitter datasets corresponding to several recent and sequential time periods.

Fig. 4 depicts the proposed idea. For example, if **I = 6,** this indicates six timely sequential snapshots of the users' data and metadata (such as the last six years, months, weeks, etc.) which are incorporated to measure the credibility of users in each time period, and the overall credibility **TC**. The users' credibility values in all time periods are indexed sequentially starting from the oldest time period.

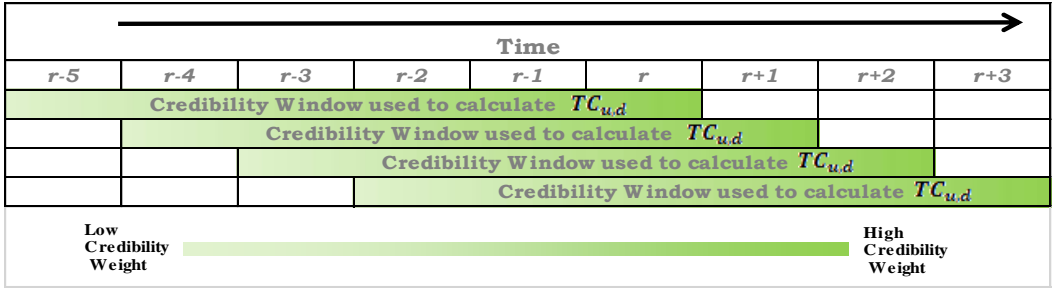
Fig. 4. Credibility Window

The threshold ($I$) is used to facilitate the credibility analysis by focusing on recent users' content in the past ($I$) time periods. Furthermore, it is more efficient to measure the credibility of users based on their current and recent behavior. This is logical since the user's interest(s) could change, and their knowledge evolves over time. Hence, the user's older content and metadata should be considered as a legacy chunk of data and therefore should not be incorporated in the credibility analysis.

### 5.2.5 Scale credibility values

The last step in our approach is to use a scale as a measurement system in order to interpret the numeric values resulting from the evaluation approach and convert them to a meaningful presentation. Thus, we customized one of the most popular scale systems which uses a 7-level trustworthiness scale [31]. This trustworthiness measure helps to rate trust by numerically quantifying the trust values and qualifying the trust levels numerically. Table 3 shows the seven levels of trustworthiness determined by this method.

Table 3: Seven levels of trustworthiness [31]

| Trustworthiness Level | Semantics (Linguistic Definitions) | Trustworthiness Value (User defined) | Visual Representation |
|---|---|---|---|
| **Level -1** | New User | $TC'_{u,d} = $ "" | Not displayed |
| **Level 0** | Very Untrustworthy User | $TC'_{u,d} = 0$ | Not displayed |
| **Level 1** | Untrustworthy User | $0 < TC'_{u,d} \leq 1$ | From ★ |
| **Level 2** | Partially Trustworthy User | $1 < TC'_{u,d} \leq 2$ | From ★ to ★★ |
| **Level 3** | Largely Trustworthy User | $2 < TC'_{u,d} \leq 3$ | From ★★ to ★★★ |
| **Level 4** | Trustworthy User | $3 < TC'_{u,d} \leq 4$ | From ★★★ to ★★★★ |
| **Level 5** | Very Trustworthy User | $4 < TC'_{u,d} \leq 5$ | From ★★★★ to ★★★★★ |

Entries of $TC$ are scaled to values between "0" and "5" as follows:

$$TC'_{u,d} = \frac{(TC_{u,d} - min(TC_{*,d})) * 5}{max(TC_{*,d}) - min(TC_{*,d})} \quad (15)$$

The next sections demonstrate the implementation of the time-aware credibility mechanism and provide an evaluation metric for the proposed credibility approach.

## 6. EXPERIMENTAL RESUTLS

To evaluate the effectivenece of CredSaT, several experiments are conducted as follows; (i)we benchmark our approach against state-of-art baseline models in indicating the highly trustworthy, domain-based influencers; (ii)we provide a closer look at the highest trustworthy users achieved in four domains of knowledge; (iii)finally, we show the capability of CredSaT to infer anomalous users.

### 6.1 Subsets selection and experiments settings

The cleansed dataset is divided into six chunks starting at Nov-2014 and ending in Apr-2015, i.e. $I = 6$, see Eq. (14), where each chunk is comprised of the data and metadata of each particular month. These chunks embody the chronologically sequential snapshots indicating the recent user's activity amongst the crawled dataset. Table 4 shows the total count of *users*, *tweets* and their *replies* for the determined time. The number of users shown in Table 4 (i.e. 6,066) represents the total distinct number of users who posted tweets in one or more of the determined months. The remaining users posted their tweets before that, although they have been inactive in twitter recently. This signifies the importance of studying users' content temporally.

Table 4: Total monthly count of users, tweets and replies

| *Month* | Nov-14 | Dec-14 | Jan-15 | Feb-15 | Mar-15 | Apr-15 | Total |
|---|---|---|---|---|---|---|---|
| *#Users* | 4,531 | 4,596 | 4,718 | 4,690 | 4,388 | 4,309 | 6,066 |
| *#Tweets* | 119,847 | 123,304 | 145,768 | 147,145 | 144,529 | 137,567 | 818,160 |
| *#Replies* | 55,949 | 58,956 | 76,561 | 73,867 | 70,135 | 61,352 | 396,820 |

The aforementioned set of equations (1) to (14) have been implemented for the datasets of each selected month. The value of $\rho$ indicated in Eq. (5) is set to "2" experimentally as it represents the monthly threshold value. Significant adjustments introduced in Eq. (13), $\alpha, \beta, \gamma, \delta, \theta$ and $\vartheta$, are empirically set to (0.2, 0.2, 0.2, 0.1, and 0.2) respectively. Function $w(k)$ indicated in Eq. (14) is defined as $w(k) = k = 6$, this implies that $C_{u,d}^k$ will be assigned a value weight equal to the associated $k = 6$ value. Hence, the highest $k$ value is assigned to the most recent dataset; conversely, the older the dataset, the lower is the assigned $k$ value. The time-aware, domain-based user's credibility matrix $TC$ is calculated for all AlchemyAPI's 23 domains of knowledge for every month of the credibility window (I), where I = 6. The time-aware, domain-based normalized credibility matrix $TC'$ is calculated. This matrix includes a ranked list of users in each particular domain.

The top users in each domain comprise the trustworthy, and very trustworthy users in that domain. Those users embody the influential users in each domain of knowledge. Although the domain of knowledge for some influencers is explicitly indicated in their twitter's bio, the domain of interest is a tacit knowledge for other domain-based, highly trustworthy users. Our approach determines those users, assigns them trustworthiness values, and places them at various levels of trustworthiness.

## 6.2 Discovery of domain-based influencers - Baseline Comparison

We benchmark CredSaT against a set of evaluation techniques over a curated labelled dataset. This dataset contains four domains ("*Computing and Technology*", "*Sports*", "*Education*", and "*Arts and Entertainment*"), and a set of "20" selected influential users in each domain. The list of influential users are selected by carefully examining their tweets, and collected metadata (bio information, #followers, #friends, etc.), thus choosing the list of users who have shown a noticeable and capacious interest in the selected domains consolidated with the figures captured from their metadata. The list of methods incorporated in the conducted comparison includes:

- *Twitterrank*[20]: aims to find topic-based influential twitterers incorporating LDA statistical model for topic distillation, and topic-sensitive PageRank for credibility propagation. As topics of users are identified in TwitterRank based on the words' distribution of their tweets, the high-level topics classifications are inadequate and inferior[24]. Therefore, the topics identified by Twitterrank(*namely LDA*) may not match our high-level domains which are identified incorporating ontology and semantic analysis facilitated by AlchemyAPI. To establish a common ground to conduct the comparison, several trials of Twitterrank are reported over our collected dataset to find closely matching topics to the four domains of knowledge, and to infer the top influential users of each topic accordingly. We adopt and customize the python implementation of Twitterank[2].

- *High In-degree:* measures the influence of twitter by studying the number of followers. This feature is incorporated by several service providers[20].

- *High domain-based key attributes:* this method extracts five lists indicating the key attributes explained in this paper and as summarized in Eq. (13). Each list comprises the set of users obtained the highest domain based values in each corresponding key attribute.

*Evaluation Metric*: The performance of finding domain-based influencers of each method is measured based on the obtained *Precision*, *Recall, F-score and nDCG*. Let $HC_d$ presents the set of influencers of domain $d$ as indicated in the curated dataset, $HR_d^Q$ embodies the top $Q$ users of domain $d$ retrieved by each incorporated method. The evaluation metrics can be calculated as the following: *Precision[1]*: it measures the ratio between the numbers of correct retrieved domain-based influential users to the number of top-$Q$ returned users by the method. *Precision* is assigned a number "1" as this metric will be utilized later for a different purpose in a different experiment (see section 6.5). *Recall:* indicates the ratio between the numbers of correct retrieved domain-based influential users to the actual number of domain-based influential users identified in the curated dataset. *F-score*: is used to provide the trade-off between *Precision* and *Recall*. *Normalized Discounted Cumulative Gain (NDGC)*: measures the performance of the model incorporating graded relevance metric. The later metric is adopted in this experiment to provide a fine grain evaluation analysis. This is through assessing the retrieved user in each method by a scale of four relevance degrees; *highly influential, influential, somehow influential, not an influential.* These metrics can be defined as follows:

$$Precision_{d_Q}^1 = \frac{|HC_d \cap HR_d^Q|}{|HC_d|} \qquad (16)$$

$$Recall_{d_Q} = \frac{|HC_d \cap HR_d^Q|}{|HR_d^Q|} \qquad (17)$$

$$F-score_{d_Q} = \frac{2*Precision_d^1*Recall_d}{Precision_d^1+Recall_d} \qquad (18)$$

$$nDCG_{d_Q} = \frac{DCG_{d_Q}}{(ideal)DCG_{d_Q}}, \text{ where } DCG_{d_Q} = \sum_{i=1}^{Q} \frac{2^{rel_{d_i}}-1}{\log_2(i+1)} \qquad (19)$$

---

[2] bit.ly/2qcSQ01bi

The conducted experiment retrieves the top 150 influencers in each domain $d$ for each model (i.e. Q=150). $Precision_d^1$, $Recall_d$, $Fscore_d$ and $nDCG_d$ are calculated for each domain, and the average is computed for each metric in all domains. Table 5 shows the performance of each model.

Table 5: Evaluation of domain-based Influential Retrieval

| Baseline Model | Precision[1] | Recall | F-Measure | nDCG |
|---|---|---|---|---|
| **CredSaT** | **0.95** | **0.75** | **0.85** | **0.93** |
| TwitterRank | 0.73 | 0.65 | 0.69 | 0.86 |
| High In-degree | 0.83 | 0.35 | 0.49 | 0.75 |
| High W' | 0.76 | 0.59 | 0.66 | 0.86 |
| High P' | 0.51 | 0.31 | 0.39 | 0.36 |
| High L' | 0.83 | 0.4 | 0.54 | 0.75 |
| High R' | 0.90 | 0.62 | 0.77 | 0.68 |
| High S' | 0.85 | 0.66 | 0.74 | 0.69 |

The figures in Table 5 indicate that our model outperforms other methods in all metrics. It is intuitive that CredSaT overshadows Twitterrank task of inferring influential users. This is because the mechanism followed by CredSaT considers several focal dimensions which are neglected by Twitterrank such as mentoring users' credibility over time, sentiments analysis of the tweets replies, etc. CredSaT as a comprehensive framework performs better than harnessing each key attributes separately to measure the users influence. For example, although the weight assigned to each user in each domain is important to address the interests of OSNs' user, this is insufficient to rank users based merely on their domains of interest. Likewise, obtaining a high number of followers does not definitely imply an influence in all domains of knowledge; yet, these number of followers might be attained due to the importance of the user in a certain domain(s). Hence, it is essential to possess an understanding of the user's interests in all domains which includes the interactions-based attributes of users in OSNs. This involves analysing the user's content, studying the overall followers' interest in the user's content, followers' sentiments toward the user, etc.

## 6.3 Highly domain-based trustworthy users

Fig. 5 shows a closer look to the CredSaT's top five trustworthy users in each selected domain of the crawled dataset. The results shown in these charts are broadly acceptable. In the "*Computing and Technology*" domain **@edithyeung, @wolf_gregor, @johnjwall, @commadelimited** and **@JeremyKendall** attained the highest positions. **@edithyeung** for example obtained a domain-based "*Very Trustworthy*" level. This is because **@edithyeung** shows a continuing interest in IT aspects in most of the posted tweets and links in twitter. Moreover, a recent visit[3] to the user profile exhibited more than 300% increase in the number of followers since this metadata was crawled during the dataset acquisition phase. This is supported by the high number of positive replies, retweets, and favourites. This applies also to the other top four users in the "Computing and Technology" domain.

**@SpnMaisieDaisy** obtained a "very trustworthy" level in "Art and Entertainment". This user often tweeted about movies and TV series, and the metadata shows that other users pay particular attention to his "Art and Entertainment"-related tweets. **@SpnMaisieDaisy** has maintained his leading position in almost every month, which indicates his continuous interest in this domain. In the "sports" domain, **@nwipreps,** presents a platform to distribute tweets about many kind of sports. This user keeps the followers updated on all sports-related news. With the highest values in the number of likes, and retweets in the sports domain, **@nwipreps** deserves to be placed in this position. In the "Law, govt, and politics" domain, the top five users in general tweeted about topics related to Low, government or politics. For example, **@englishvoice** is the official twitter account for the English Democrats, the nationalist political party in England. It is reasonable to expect that their twitter account would achieve a five-star ranking because this account is dedicated to discussing political topics, which is supported by their followers. Apart from his interest in politics-related news, **@IvorCrotty** indicates in his bio that he is the head of a social media extension "**@rt_com**" for "Russia Today": a Russian government-funded television network. Thus, **@IvorCrotty** has maintained his dominant position in the "Law, govt, and politics" domain.

---

[3] https://twitter.com/edithyeung, Visited in 30/04/2017

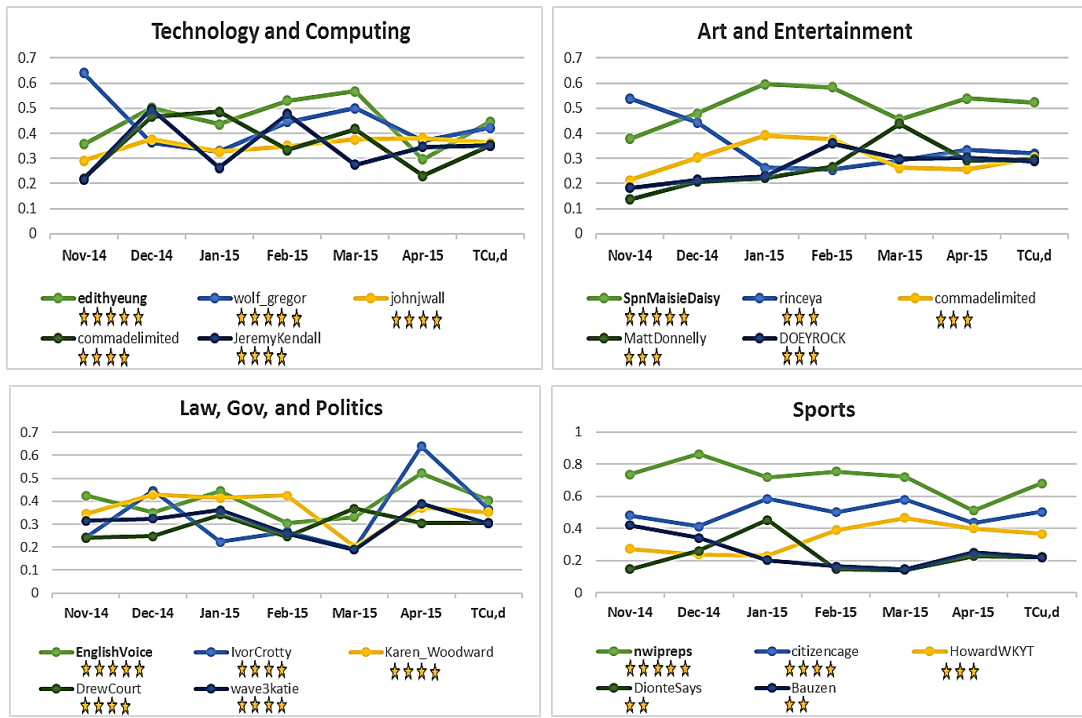

Fig. 5. Highest $TC_{u,d}$ values in four selected domains

### 6.4 Discovery of Anomalous users

In addition to all the levels of trustworthy users, $TC_{u,d}$ comprises a wide range of domain-based "*Untrustworthy*" and "*Very Untrustworthy*" users; thus, it is highly likely that this indicates that spammers, or other illegitimate user categories are amongst them. In order to capture these categories more easily, we apply two criteria to narrow our research:

- Selecting a set of users who have been placed at a "Very Untrustworthy" level (i.e $TC_{u,*} = 0$) in ALL 23 domains, and achieved the lowest values in Tweets Similarity Penalty ($Twt\_Sim$)(i.e strong similarity of tweets).

- Selecting a set of users who have been placed at a "Very Untrustworthy" level (i.e $TC_{u,*} = 0$) in ALL 23 domains and achieved the lowest values in URL Similarity Penalty ($URL\_Sim$)(i.e. strong similarity of URLs).

The results of the above criteria are compared with a set of retrieved users based on the following criteria:

- *Low In-degree*: selects users who obtained the least number of followers.

- *Anomaly detection toolkit of Graphlab™[4]:* This machine learning based module indicates the data items/points which are different from other data items. It assigns an anomaly score of value between "0" to "∞", where the higher the score, the more likely the data item is anomalous. All users with the following features were passed to this toolkit; ***#DistinctWords***, ***#Words, Twt_Sim***, and the ***TC*** values of all 23 domains. The users who achieved the highest score; i.e. detected anomalies, are used in this benchmark.

---
[4] https://turi.com/products/create

The examination process was conducted manually by reading all the crawled tweets of each user in each criterion's set, and labeling each user with one of two main categories (i) *Normal* users: are those legitimate users whose tweeting behavior is normal; (ii) *Anomalous* users: are those who utilize the Twitter platform for scamming, spamming, and other anomalous activities. The precision evaluation metric is computed as follows:

$$Precision^2 = \frac{Number\ of\ perceived\ anomalous\ users}{Number\ of\ retrieved\ users} \quad (20)$$

Fig. 6 presents the retrieval precision of the top-K at 10, 20, 30, 40, 50 and the Average Precision. As depicted in this figure, the experiments conducted on the retrieved users of each criterion verify the effectiveness of our approach to discover anomalous users. For example, the first 10 users retrieved by enquiring users who obtained zero credibility value in all domains, along with their tweets' similarities are the highest (i.e. lowest $Twt\_Sim$ values), were all exhibiting anomalous behavior. However, only one user of the first 10 retrieved users, whose in-degree features are the lowest, was anomalous. Although the anomalous users discovered using the criterion "*Very Untrustworthy with Low Twt_Sim*" are relatively similar to those detected using "*Graphlab-anomaly_detection*" module, the average precision accumulated using our approach is promising for building anomalies detection frameworks consolidated with the features proposed in this paper. This will be investigated further in our future work.

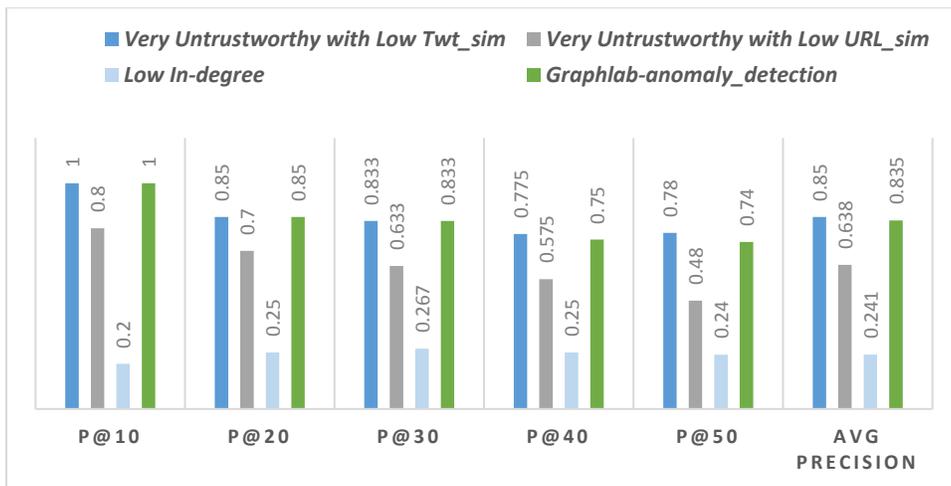

Fig. 6 Evaluation of Anomalous Retrieval (*precesion²*)

# 6   DISCUSSION, LIMITATIONS AND FUTURE RESEARCH

CredSaT is an ongoing project; the proposals put forward in this paper have evolved since our preliminary work [9]. The main objective is to provide a methodology for BSD that incorporates trust and semantic analysis. For future work, several enhancements will be implemented to consolidate the proposed approach:

- CredSaT will be improved to handle two additional BD features: Variety through the importation of more data sources; and Velocity through the addition of a new module to measure the credibility of the new content in real time (i.e. assign a credibility value to a new user's tweet).
- AlchemyAPI has been used in this paper as the sole semantics provider. Although this service provider is supported by IBM, a prestigious software company, the resultant semantics should be enhanced further by utilizing an ontology-based approach. In particular, domain knowledge is captured in ontologies which are then used to enrich the semantics of tweets provided with specific semantic conceptual representation of entities that appear in the tweets.
- A new graph-based model will be created to propagate the users' credibility throughout the entire network. Hence, an enhanced version of Twitterrank [20] is anticipated that takes into consideration the semantics of the textual content and the temporal factor.
- An anomaly detection approach will be developed that incorporates machine learning and an advanced list of features.

# 7   CONCLUSION

This paper presents the CredSaT (*Credibility incorporating Semantic analysis and Temporal factor*): a domain-based credibility framework incorporating semantic analysis and the temporal factor to measure and rank the credibility of users in BSD. We incorporate the BD value chain presented by Hu et al. [25] which covers the life cycle of BD. CredSaT addresses four main BD features: *Veracity* through data trustworthiness, data certainty and reliable data store; *Volume* through BD storage cluster; *Variability* by incorporating semantic analysis; and *Value* by creating a comprehensive framework to measure the credibility of users in BSD.

The core of the credibility module of CredSaT is constructed based on three main dimensions: (i) distinguishing users of the various domains of knowledge; (ii) a novel metric incorporating a list of fine-grained key attributes is harnessed to create the feature-based ranking model; and (iii) the temporal factor is used to study the users' behaviour over time and reflect this behaviour by means of their domain-based credibility values.

The experiments conducted to evaluate this approach validate the applicability and effectiveness of determining highly domain-based trustworthy users, as well as capturing spammers and other low trustworthy users.

## REFERENCES

1. Chen, H., R.H. Chiang, and V.C. Storey, *Business Intelligence and Analytics: From Big Data to Big Impact*. MIS Quarterly, 2012. **36**(4).
2. Emani, C.K., N. Cullot, and C. Nicolle, *Understandable Big Data: A survey*. Computer Science Review, 2015. **17**: p. 70-81.
3. Hitzler, P. and K. Janowicz, *Linked Data, Big Data, and the 4th Paradigm*. Semantic Web, 2013. **4**(3): p. 233-235.
4. Kumar, A. and T.M. Sebastian, *Sentiment analysis on twitter*. IJCSI International Journal of Computer Science Issues, 2012. **9**(3): p. 372-378.
5. Zhang, B., et al., *A trust-based sentiment delivering calculation method in microblog*. International Journal of Services Technology and Management, 2015. **21**(4-6): p. 185-198.
6. Bae, Y. and H. Lee, *Sentiment analysis of Twitter audiences: Measuring the positive or negative influence of popular twitterers*. Journal of the American Society for Information Science and Technology, 2012. **63**(12): p. 2521-2535.
7. Kawabe, T., et al. *Tweet credibility analysis evaluation by improving sentiment dictionary*. in *Evolutionary Computation (CEC), 2015 IEEE Congress on*. 2015. IEEE.
8. Demchenko, Y., et al. *Addressing big data issues in scientific data infrastructure*. in *Collaboration Technologies and Systems (CTS), 2013 International Conference on*. 2013. IEEE.
9. Abu-Salih, B., P. Wongthongtham, and D. Zhu. *A Preliminary Approach to Domain-Based Evaluation of Users' Trustworthiness in Online Social Networks*. in *2015 IEEE International Congress on Big Data*. 2015. IEEE.
10. Cha, M., et al., *Measuring User Influence in Twitter: The Million Follower Fallacy*. ICWSM, 2010. **10**: p. 10-17.
11. Silva, A., et al. *ProfileRank: finding relevant content and influential users based on information diffusion*. in *Proceedings of the 7th Workshop on Social Network Mining and Analysis*. 2013. ACM.
12. Jiang, W., G. Wang, and J. Wu, *Generating trusted graphs for trust evaluation in online social networks*. Future generation computer systems, 2014. **31**: p. 48-58.
13. Zhao, L., et al., *A topic-focused trust model for twitter*. Computer Communications, 2015.
14. Brown, P.E. and J. Feng. *Measuring user influence on twitter using modified k-shell decomposition*. in *Fifth International AAAI Conference on Weblogs and Social Media*. 2011.
15. Zhu, Z., J. Su, and L. Kong, *Measuring influence in online social network based on the user-content bipartite graph*. Computers in Human Behavior, 2015. **52**: p. 184-189.
16. Embar, V.R., et al., *Online Topic-based Social Influence Analysis for the Wimbledon Championships*, in *Proceedings of the 21th ACM SIGKDD International Conference on Knowledge Discovery and Data Mining*. 2015, ACM: Sydney, NSW, Australia. p. 1759-1768.
17. Zhou, J., Y. Zhang, and J. Cheng, *Preference-based mining of top-K influential nodes in social networks*. Future Generation Computer Systems, 2014. **31**: p. 40-47.
18. Wei, W., et al., *Learning to Find Topic Experts in Twitter via Different Relations*. IEEE Transactions on Knowledge and Data Engineering, 2016. **28**(7): p. 1764-1778.
19. Pal, A., et al. *Discovery of topical authorities in instagram*. in *Proceedings of the 25th International Conference on World Wide Web*. 2016. International World Wide Web Conferences Steering Committee.
20. Weng, J., et al. *Twitterrank: finding topic-sensitive influential twitterers*. in *Proceedings of the third ACM international conference on Web search and data mining*. 2010. ACM.
21. Chandrasekaran, B., J.R. Josephson, and V.R. Benjamins, *What are ontologies, and why do we need them?* IEEE Intelligent systems, 1999. **14**(1): p. 20-26.

22. Alahmadi, D.H. and X.-J. Zeng, *ISTS: Implicit social trust and sentiment based approach to recommender systems.* Expert Systems with Applications, 2015. **42**(22): p. 8840-8849.
23. AlRubaian, M., et al. *A Multistage Credibility Analysis Model for Microblogs*. in *Proceedings of the 2015 IEEE/ACM International Conference on Advances in Social Networks Analysis and Mining 2015*. 2015. ACM.
24. Michelson, M. and S.A. Macskassy. *Discovering users' topics of interest on twitter: a first look*. in *Proceedings of the fourth workshop on Analytics for noisy unstructured text data*. 2010. ACM.
25. Han, H., et al., *Toward Scalable Systems for Big Data Analytics: A Technology Tutorial.* Access, IEEE, 2014. **2**: p. 652-687.
26. Makice, K., *Twitter API: Up and running: Learn how to build applications with the Twitter API*. 2009: " O'Reilly Media, Inc.".
27. Akcora, C.G., et al., *Detecting anomalies in social network data consumption.* Social Network Analysis and Mining, 2014. **4**(1): p. 1-16.
28. Saravanakumar, M. and T. SuganthaLakshmi, *Social media marketing.* Life Science Journal, 2012. **9**(4): p. 4444-4451.
29. Gentner, D. and A.L. Stevens, *Mental models*. 1983, Hillsdale, N.J: L. Erlbaum Associates.
30. Wang, A.H. *Don't follow me: Spam detection in Twitter*. in *Security and Cryptography (SECRYPT), Proceedings of the 2010 International Conference on*. 2010.
31. Chang, E., F. Hussain, and T. Dillon, *Trust and reputation for service-oriented environments: technologies for building business intelligence and consumer confidence*. 2006: John Wiley & Sons.